\documentclass[a4paper,12pt]{article}
\usepackage{amssymb,amsmath,amsfonts}
\usepackage{fullpage}
\usepackage{graphicx}
\usepackage[latin1]{inputenc}
\begin{document}
\title{The distribution of Pearson residuals in generalized linear models}
\date{}
\author{By Gauss M. Cordeiro$^{a,}$\footnote{Corresponding author. E-mail: gausscordeiro@uol.com.br}~ and Alexandre B. Simas$^{b,}$\footnote{E-mail: alesimas@impa.br}\\\\
\centerline{\small{
$^a$Departamento de Estat\' \i stica e Inform\' atica,
Universidade Federal Rural de Pernambuco,
}}\\
\centerline{\small{
Rua Dom Manoel de Medeiros s/n, Dois Irm\~ aos,
52171-900 Recife-PE, Brasil}}\\
\centerline{\small{
$^b$Associa\c{c}\~ao Instituto Nacional de Matem\'atica Pura e Aplicada, IMPA,}}\\
\centerline{\small{
Estrada D. Castorina, 110, Jd. Bot\^anico, 22460-320, Rio de Janeiro-RJ, Brasil}}}

\maketitle
\begin{abstract}
In general, the distribution of residuals cannot be obtained
explicitly. We give an asymptotic formula for the density of Pearson
residuals in continuous generalized linear models corrected to order
$n^{-1}$, where $n$ is the sample size. We define corrected Pearson
residuals for these models that, to this order of approximation,
have exactly the same distribution of the true Pearson residuals.
Applications for important generalized linear models are provided
and simulation results for a gamma model illustrate the usefulness
of the corrected Pearson residuals.
\end{abstract}

\section{Introduction}
The residuals carry important information concerning the
appropriateness of assumptions that underlie statistical models, and
thereby play an important role in checking model adequacy. They are
used to identify discrepancies between models and data, so it is
natural to base residuals on the contributions made by individual
observations to measures of model fit. The use of residuals for
assessing the adequacy of fitted regression models is nowadays
commonplace due to the widespread availability of statistical
software, many of which are capable of displaying residuals and
diagnostic plots, at least for the more commonly used models. Beyond
special models, relatively little is known about asymptotic
properties of residuals in general regression models. There is a
clear need to study second-order asymptotic properties of
appropriate residuals to be used for diagnostic purposes in
nonlinear regression models.

The unified theory of generalized linear models (GLMs), including a
general algorithm for computing the maximum likelihood estimates
(MLEs) is extremely important for analysis of real data. In these
models, the random variables $Y_1,\ldots,Y_n$ are assumed
independent and each $Y_i$ has a density function in the linear
exponential family
\begin{equation}\label{exp}
\pi(y; \theta_i,\phi) = \exp[\phi\{y\theta_i - b(\theta_i) \} + c(y,\phi)],
\end{equation}
where $b(\cdot)$ and $c(\cdot,\cdot)$ are known appropriate
functions. We assume $Y$ continuous and $\pi$ a probability density
function with respect to Lebesgue measure and that the precision
parameter $\phi=\sigma^{-2}$, $\sigma^{2}$ is the so-called
dispersion parameter, is the same for all observations, although
possibly unknown. We do not consider the discrete distributions in
the form ($\ref{exp}$) such as Poisson, binomial and negative
binomial. For two-parameter full exponential family distributions
with canonical parameters $\phi$ and $\phi \theta$, the
decomposition $c(y,\phi)=\phi a(y)+ d_1(y)+d_2(\phi)$ holds. The
mean and variance of $Y_i$ are, respectively, $E(Y_i) = \mu_i = d
b(\theta_i)/d\theta_i$ and ${\rm Var}(Y_i) = \phi^{-1} V_i$, where
$V = d\mu/d\theta$ is the variance function. For gamma models, the
dispersion parameter $\sigma^2$ is the reciprocal of the index,
whereas for normal and inverse Gaussian models, $\sigma^2$ is the
variance and ${\rm Var}(Y_i)/E(Y_i)^3$, respectively. The parameter
$\theta = \int V^{-1} d\mu = q(\mu)$ is a known one-to-one function
of $\mu$. A linear exponential family is characterized by its
variance function, which plays a key role in estimation.

A GLM is defined by the family of distributions (\ref{exp}) and the
systematic component $g(\mu) = \eta= X\beta$, where $g(\cdot)$ is a
known one-to-one continuously twice-differentiable function, $X$ is
a specified $n\times p$ model matrix of full rank $p<n$ and $\beta =
(\beta_1,\ldots,\beta_p)^T$ is a set of unknown linear parameters to
be estimated. Let $\hat\beta$ be the MLE of $\beta$.

Residuals in GLMs were first discussed by Pregibon (1981), though
ostensibly concerned with logistic regression models, Williams
(1984, 1987) and Pierce and Schafer (1986). McCullagh and Nelder
(1989) provided a survey of GLMs with substantial attention to
definition of residuals. Pearson residuals are the most commonly
used measures of overall fit for GLMs and are defined by $R_i =
(Y_i - \hat{\mu_i})/\hat{V_i}^{1/2}$, where $\hat{\mu_i}$ and
$\hat{V_i}$ are respectively the fitted mean and fitted variance
function of $Y_i$. In this paper we consider only Pearson residuals
appropriate to our particular asymptotic aims when the sample size
$n\rightarrow \infty$. Cordeiro (2004) obtained matrix formulae for
the expectations, variances and covariances of these residuals and
defined adjusted Pearson residuals having zero mean and unit
variance to order $n^{-1}$. Pearson residuals defined by
Cordeiro (2004) are proportional to $\sqrt{\phi}$, although we are
considering here $R_i$ as usual without the precision parameter
$\phi$. While Cordeiro's adjusted Pearson residuals do correct the
residuals for equal mean and variance, the distribution of these
residuals is not equal to the distribution of the true Pearson
residuals to order $n^{-1}$.

Further, Cordeiro and Paula (1989) introduced the class of
exponential family nonlinear models (EFNLMs) which extend the GLMs. 
Later, Wei (1998) gave a comprehensive introduction to these
models. Recently, Simas and Cordeiro (2008) generalized Cordeiro's
(2004) results by obtaining matrix formulae of the ${\cal O}(n^{-1})$ 
expectations, variances and covariances of Pearson residuals in EFNLMs.

In a general setup, the distribution of residuals usually differ
from the distribution of the true residuals by terms of order
$n^{-1}$. Cox and Snell (1968) discussed a general definition of
residuals, applicable to a wide range of models, and obtained useful
expressions to this order for their first two moments. Loynes (1969)
derived, under some regularity conditions, and again to order
$n^{-1}$, the asymptotic expansion for the density function of 
Cox and Snell's residuals, and then defined corrected residuals
having the same distribution as the random variables which they are
effectively estimating. In all but the simplest situations, the use
of the results by Cox and Snell and Loynes will require a
considerable amount of tedious algebra. Our chief goal is to obtain
an explicit formula for the density of Pearson residuals to
order $n^{-1}$ which holds for all continuous GLMs.

In Section 2 we give a summary of key results from Loynes (1969)
applied to Pearson residuals in GLMs. The density of Pearson
residuals in these models corrected to order $n^{-1}$ is presented
in Section 3. We provide in Section 4 applications to some common
mo\-dels. In Section 5 we compare the corrected residuals with the
adjusted residuals proposed by Cordeiro (2004). We present in
Section 6 simulation studies to assess the adequacy of the
approximations for a gamma model with log link. Some
concluding remarks are given in Section 7. Finally, in the
Appendix, we give a more rigorous proof of the general results
discussed by Loynes (1969).

\section{Conditional moments of Pearson residuals}

The $i$th contribution for the score function from the observation
$Y_i$ follows from (\ref{exp})
$$U_r^{(i)} = \frac{\partial l_i}{\partial\beta_r} = \phi V_i^{-1/2}w_i^{1/2}(Y_i - \mu_i) x_{ir},$$
where $w = V^{-1} \mu'^2 $ is the weight function and from now on
the dashes indicate derivatives with respect to $\eta$. Let
$\varepsilon_i = V_i^{-1/2}(Y_i - \mu_i)$ be the true Pearson
residual corresponding to the Pearson residual
$R_i=\hat{V_i}^{-1/2}(Y_i-\hat{\mu}_i)$. Suppose we write the
Pearson residual as $R_i=\varepsilon_i+\delta_i$. We can write the
following conditional moments given $\varepsilon_i = x$ to order
$n^{-1}$ (Loynes, 1969)
$${\rm Cov}(\hat{\beta}_r,\hat{\beta}_s\mid \varepsilon_i = x)=-\kappa^{rs},$$
\begin{equation}\label{biascond}
b_s^{(i)}(x) = E(\hat{\beta}_s-\beta_s\mid\varepsilon_i=x)=B(\hat{\beta}_s)-\sum_{r=1}^p\kappa^{sr}U_r^{(i)}(x),
\end{equation}
where $-\kappa^{sr}$ is the $(s,r)$th element of the inverse 
information matrix $K^{-1}$ for $\beta$, $B(\hat{\beta}_s)$ is the
${\cal O}(n^{-1})$ bias of $\hat{\beta}_s$ and $U_r^{(i)}(x) =
E(U_r^{(i)}\mid \varepsilon_i = x)$ is the conditioned score
function. The mean and variance of the asymptotic distribution of
$\delta_i$, given $\varepsilon_i = x$, are to order $n^{-1}$
\begin{equation}\label{meancond}
\theta_x^{(i)}=E(\delta_i\mid\varepsilon_i=x)=\sum_{r=1}^p H_r^{(i)}(x) b_r^{(i)}(x)-\frac{1}{2}\sum_{r,s}^p H_{rs}^{(i)}(x)\kappa^{rs},
\end{equation}
\begin{equation}\label{varcond}
\phi_x^{(i)^2}={\rm Var}(\delta_i\mid\varepsilon_i=x)=-\sum_{r,s=1}^p H_r^{(i)}(x)H_s^{(i)}(x)\kappa^{rs},
\end{equation}
where $H_r^{(i)} = \partial\varepsilon_i/\partial\beta_r$, $H_{rs}^{(i)} = \partial^2\varepsilon_i/\partial\beta_r\partial\beta_s$,
$H_r^{(i)}(x) = E(H_r^{(i)}\mid \varepsilon_i=x)$ and $H_{rs}^{(i)}(x)=E(H_{rs}^{(i)}\mid \varepsilon_i = x)$.
We obtain by simple differentiation
$$H_r^{(i)} = \{-V_i^{-1/2}\mu_i'-\frac{1}{2} V_i^{-3/2} V_i^{(1)}\mu_i'(Y_i - \mu_i)\}\,x_{ir}$$
and
\begin{eqnarray*}H_{rs}^{(i)} &=&\{-V_i^{-1/2}\mu_i''+ V_i^{-3/2} V_i^{(1)}\mu_i'^2+\frac{3}{4}V_i^{-5/2}V_i^{(1)^2}\mu'^2(Y_i-\mu_i)\\
&-& \frac{1}{2} V_i^{-3/2}V_i^{(2)}\mu_i'^2(Y_i-\mu_i)-\frac{1}{2} V_i^{-3/2}V_i^{(1)}\mu_i'(Y_i-\mu_i)\}\,x_{ir}x_{is}.
\end{eqnarray*}
Conditioning on $\varepsilon_i = x$ leads to $H_r^{(i)}(x) = e_i (x) x_{ir}$ and
$H_{rs}^{(i)}(x)=h_i(x)x_{ir}x_{is},$ where
\begin{equation}\label{efunction}
e_i(x) = -V_i^{-1/2}\mu_i'-\frac{1}{2}V_i^{-1}V_i^{(1)}\mu_i' \, x
\end{equation}
and
\begin{equation}\label{hfunction}
h_i(x) = -V_i^{-1/2}\mu_i''+ V_i^{-3/2}V_i^{(1)}\mu_i'^2 + \frac{1}{4}\{(3 V_i^{-2} V_i^{(1)^2}
- 2 V_i^{-1}V_i^{(2)}) \mu_i'^2 - 2 V_i^{-1}V_i^{(1)}\mu_i''\}\,{x}.
\end{equation}
For canonical models ($\theta = \eta$), (\ref{efunction}) and (\ref{hfunction}) become
$$e_i(x)=-V_i^{1/2}-\frac{V_i^{(1)}}{2} x \hbox{~ ~ and ~ ~}  h_i(x) = \frac{1}{4}(V_i^{(1)^2} - 2 V_i V_i^{(2)}) x.$$
Conditioning the score function $U_r^{(i)} = \phi V_i^{-1/2} w_i^{1/2} (Y_i - \mu_i)x_{ir}$
on $\varepsilon_i = x$, yields $U_r^{(i)}(x)=\phi \, w_i^{1/2} \, x_{ir} \, x $, and then
using ($\ref{biascond}$) we find
\begin{eqnarray*}
b_s^{(i)}(x) &=& B(\hat{\beta}_s)+\phi \,w_i^{1/2} \,\tau_s^T K^{-1} X^T\gamma_i\,x,
\end{eqnarray*}
where $K^{-1} = \phi^{-1}(X^T W X)^{-1}$, $W = {\rm diag}\{w_i\}$ is
the diagonal matrix of weights, $\tau_s$ is a $p$-vector with the
$s$th element equal to one and all other elements equal to zero and
$\gamma_i$ is an $n$-vector with one in the $i$th position and zeros
elsewhere. Defining $M =\{m_{si}\}=(X^T W X)^{-1} X^T$, we can
easily verify that
$$b_s^{(i)}(x) = B(\hat{\beta}_s)+w_i^{1/2}\,m_{si}\,x.$$
Cordeiro and McCullagh (1991) showed that the $n^{-1}$ bias of
$\hat\beta$ is given by
$$B(\hat{\beta}) = -(2\phi)^{-1}(X^T W X)^{-1}X^TZ_d \,F \,1,$$
where $F = {\rm diag}\{V_i^{-1}\mu_i'\,\mu_i''\}$, $Z = \{z_{ij}\} =
X(X^T W X)^{-1} X^T$, $Z_d = {\rm diag}\{z_{ii}\}$ is a diagonal
matrix with the diagonal elements of $Z$ and $1$ is an $n$-vector of
ones. The asymptotic covariance matrix of the MLE $\hat\eta$ of the
linear predictor is simply $\phi^{-1}\,Z$. We obtain
\begin{eqnarray*}
\sum_{r=1}^n H_r^{(i)}(x) b_r^{(i)}(x) &=& e_i(x) \{x \, w_i^{1/2}\sum_{r=1}^n m_{ri} x_{ir} + \sum_{r=1}^n B(\hat\beta_r)x_{ir}\}\\
&=&  e_i(x)\{w_i^{1/2} z_{ii}\,x + B(\hat\eta_i)\},
\end{eqnarray*}
where $B(\hat{\eta}_i)$ is the $i$th element of the ${\cal
O}(n^{-1})$ bias $B(\hat{\eta})=-(2\phi)^{-1}ZZ_d \,F\,1$ of
$\hat\eta$. The bias expression depends on the model matrix, the variance function and the first
two derivatives of the link function. Also,
\begin{eqnarray*}
-\frac{1}{2}\sum_{r,s=1}^p H_{rs}^{(i)}(x)\kappa^{rs} &=-& \frac{z_{ii}}{2\phi} h_i(x).
\end{eqnarray*}
The conditional mean $\theta_x^{(i)}$ from (\ref{meancond}) is then
a second-degree polynomial in $x$ given by
\begin{equation}\label{thetax}
\theta_x^{(i)} = \{w_i^{1/2}z_{ii} x + B(\hat{\eta}_i)\} e_i(x) + \frac{z_{ii}}{2\phi} h_i(x),
\end{equation}
where $e_i(x)$ and $h_i(x)$ are obtained from ($\ref{efunction}$) and
($\ref{hfunction}$).

We now compute the conditional variance $\phi_x^{(i)^2}$. From
($\ref{varcond}$) it follows
\begin{equation}\label{phix}
\phi_x^{(i)^2} = \frac{z_{ii}}{\phi}e_i(x)^2.
\end{equation}
Hence, $\phi_x^{(i)^2}$ is also a second-degree polynomial in $x$.

\section{The density of Pearson residuals}
A simple calculation from ($\ref{exp}$) gives the probability density function (pdf) 
of the true Pearson residual
\begin{equation}\label{densres}
f_{\varepsilon_i}(x) = \sqrt{V_i}\,{\rm exp}[\phi \{\sqrt{V_i} \,
\theta_i \,x + \mu_i \,\theta_i - b(\theta_i)\} + c(\sqrt{V_i} \,x +\mu_i,\phi)],
\end{equation}
where $\theta=q(\mu)$. Table 1 gives the densities of the true
residuals for the normal, gamma and inverse Gaussian distributions, where
$\Gamma(\cdot)$ is the gamma function.

\begin{table}[htb]
\caption{Densities of the true residuals for some distributions.}
\begin{center}
\begin{tabular}{ccc}
\hline
Distribution & Density in (1) & Density of the true residual ($f_\varepsilon(x)$)\\
\hline
Normal  & $\frac{1}{\sqrt{2\pi}\sigma}\exp\left\{-\frac{(x-\mu)^2}{2\sigma^2}\right\}$ & $\frac{1}{\sqrt{2\pi}\sigma}\exp\left(-\frac{x^2}{2\sigma^2}\right),x\in\mathbb{R}$\\
Gamma  & $\frac{(\phi x)^{\phi-1}\phi}{\Gamma(\phi)\mu^\phi}\exp(-\phi x/\mu)$ & $\frac{\{\phi (1+x)\}^{\phi-1}\phi}{\Gamma(\phi)}\exp\{-\phi (1+x)\},x>-1$\\
Inverse Gaussian & $\frac{\sqrt{\phi}}{\sqrt{2\pi x^3}}\exp\left\{-\frac{\phi (x-\mu)^2}{2\mu^2 x}\right\}$ & $\left\{\frac{\phi} {2\pi (\mu^{1/2}x+1)^3}\right\}^\frac{1}{2}\exp\left\{-\frac{\phi x^2}{2(\mu^{1/2}x+1)}\right\},x>\frac{-1}{\sqrt{\mu}}$\\
\hline
\end{tabular}
\end{center}
\end{table}
Throughout the following we assume that the standard regularity conditions of ma\-xi\-mum likelihood 
theory are satisfied. The pdf of the Pearson residual $R_i$ in continuous GLMs to order 
$n^{-1}$ follows from Loynes (1969). See, also, equation ($\ref{densloynes}$) in the Appendix. We have
\begin{equation}\label{loynesglms}
f_{R_i}(x) = f_{\varepsilon_i}(x) - \frac{d\{ f_{\varepsilon_i}(x)\theta_x^{(i)}\}} {dx} + \frac{1}{2}
\frac{d^2 \{f_{\varepsilon_i}(x) \phi_x^{(i)^2}\}}{dx^{2}},
\end{equation}
where $f_{\varepsilon_i}(x)$, $\theta_x^{(i)}$ and $\phi_x^{(i)^2}$ come from ($\ref{densres}$),
($\ref{thetax}$) and ($\ref{phix}$), respectively.

We now define corrected Pearson residuals for these models of the
form $R_i'=R_i+\rho_i (R_i)$, where $\rho(\cdot)$ is a function of order
${\cal O}(n^{-1})$ constructed in order to produce the residual
$R_i'$ with the same distribution of $\varepsilon_i$ to order
$n^{-1}$. Loynes (1969) showed (see, also, the proof given in the
Appendix) that if
\begin{equation}\label{rhoglms}
\rho_i(x) = -\theta_x^{(i)}+ \frac{1}{2 f_{\varepsilon_i}(x)}\frac{d\{f_{\varepsilon_i}(x)\phi_x^{(i)^2}\}}{dx}
\end{equation}
then $f_{R_i'}(x)=f_{\varepsilon_i}(x)$ holds to order $n^{-1}$,
i.e., the corrected residuals $R_i'$ have the same distribution of
the true residuals to this order of approximation. Combining
(\ref{phix}) with (\ref{densres}) gives
\begin{equation}\label{phidens}
\frac{1}{2f_{\varepsilon_i}(x)}\frac{d\{f_{\varepsilon_i}(x)\phi_x^{(i)^2}\}}{dx}
= \frac{z_{ii}}{\phi}\,e_i(x)\frac{d e_i(x)}{dx}+
\frac{z_{ii}}{2\phi}\,e_i(x)^2\{\phi\sqrt{V_i}\,\theta_i +
\frac{d}{dx}c(\sqrt{V_i}\,x+\mu_i,\phi)\}.
\end{equation}
Using (\ref{rhoglms}), (\ref{thetax}) and (\ref{phidens}), the correction
function turns out to be
\begin{eqnarray}
\rho_i(x) &=& e_i(x)\{-\frac{1}{2\phi}V_i^{-1}V_i^{(1)}\mu_i'\,z_{ii}-B(\hat{\eta}_i)-w_i^{1/2} z_{ii}\,x\}\nonumber\\
&-& \frac{z_{ii}}{2\phi}h_i(x)+\frac{z_{ii}}{2\phi} e_i(x)^2
\left\{\phi\sqrt{V_i}\,q(\mu_i)+\frac{d}{dx}c(\sqrt{V_i} x
+\mu_i,\phi)\right\}\label{rhopearson}.
\end{eqnarray}
Direct substitution using ($\ref{rhopearson}$) yields the corrected
Pearson residuals $R_i'$ for most models. The term $\phi^{-1}
\,z_{ii}$ in the above equation is just ${\rm Var}(\hat{\eta_i})$.
Although there are several terms in ($\ref{rhopearson}$), the correction
term is simple to be applied to any continuous model since we need only to
calculate $e_i(x), h_i(x)$ and $\frac{d}{dx} c(\sqrt{V_i}x +
\mu_i,\phi)$ from ($\ref{efunction}$), ($\ref{hfunction}$) and
($\ref{exp}$), the others terms being standard quantities in the
theory of GLMs. More generally, the corrected residuals $R_i'$
depend on the model only through the matrix $X$, the precision
parameter $\phi$, the function $c(\cdot,\cdot)$ and the variance and
link functions with their first two derivatives.

The density of the true residual for the inverse Gaussian model
given in Table 1 depends on the unknown mean $\mu$. However, we can
estimate this density using the general expression for the corrected
MLE of $\mu$, $\tilde\mu$ say, given by Cordeiro and McCullagh
(1991, formula (4.4)). The resulting estimated density is identical
to the true density except by terms of order less than $n^{-1}$ and
the results of Sections 3 and 4 could also be applied to this
distribution. To prove this, let $\tilde\mu = \mu + c/n^2.$ Then,
keeping only terms up to order $n^{-2}$, we have
$$\tilde\mu^{1/2} = \sqrt{\mu}\sqrt{1+\frac{c}{n^2\mu}} = \sqrt{\mu}\left(1+\frac{c}{2n^2\mu}\right).$$
Also,
$$(\tilde\mu^{1/2}x+1)^{-3/2}=(\sqrt{\mu}x+1)^{-3/2}\left\{1-\frac{3\,x\,c}{4n^2\sqrt{\mu}(\sqrt{\mu}x+1)}\right\}$$
and
$$\exp\left\{\frac{-\phi x^2}{2(\tilde\mu^{1/2}x+1)}\right\} = \exp\left[\frac{-\phi x^2}{2(\sqrt{\mu}x+1)}\frac{1}{\left\{1+\frac{x\,c}{2n^2\sqrt{\mu}(\sqrt{\mu}x+1)}\right\}}\right].$$
Then,
$$\exp\left\{\frac{-\phi x^2}{2(\tilde\mu^{1/2}x+1)}\right\} = \exp\left\{\frac{-\phi\,
x^2}{2(\sqrt{\mu}x+1)}\right\}\exp\left\{\frac{\phi\,x^3\,c}{4n^2\sqrt{\mu}(\sqrt{\mu}x+1)^2}\right\}.
$$
Hence,
\begin{eqnarray*}
\lefteqn{\sqrt{\frac{\phi}{2\pi}} \frac{1}{(\tilde\mu^{1/2}x+1)^{3/2}}\exp\left\{\frac{-\phi x^2}{2(\tilde\mu^{1/2}x+1)}\right\} = }\\
&\!\!\!\!\sqrt{\frac{\phi}{2\pi}}(\sqrt{\mu}x+1)^{-3/2}\left(1 - \frac{c_1}{n^2}\right)\exp\left\{\frac{-\phi x^2}{2(\sqrt{\mu}x+1)}\right\}\exp\left(\frac{c_2}{n^2}\right),\\
\end{eqnarray*}
where $c_1 = \frac{3\,x\,c}{4\sqrt{\mu}(\sqrt{\mu}x+1)}$ and $c_2 =
\frac{\phi\,x^3\,c}{4\sqrt{\mu}(\sqrt{\mu}x+1)^2}$. From this equation it is clear that
the estimated density and the true density of $\varepsilon$ are in agreement to order $n^{-1}$.

\section{Some special models}

Formula ($\ref{rhopearson}$) holds for all continuous GLMs including
the models in common use: linear models, canonical models, normal
models, gamma models and inverse Gaussian models. We now compute the
correction $\rho_i(\cdot)$ in (\ref{rhopearson}) for some important
GLMs and obtain the corrected residuals $R_i'=R_i+\rho_i(R_i)$.
Table 2 and 3 give the quantities $\mu'$, $\mu''$ and $w$ for some useful
link functions and $q(\mu)$, $V$, $w$ and $\frac{d}{dx}c(\sqrt{V}x+\mu,\phi)$
for the normal, gamma and inverse Gaussian distributions, respectively.
\begin{table}[htb]
\caption{Values of $\mu'$, $\mu''$ and $w$ for some link functions.}
\begin{center}
\begin{tabular}{ccccc}
\hline
Link function & Formula & $\mu'$ & $\mu''$ & $w$\\
\hline
Linear & $\mu=\eta$ & 1 & 0 & $V^{-1}$\\
Log & $log(\mu) = \eta$ & $\mu$ & $\mu$& $\mu^2V^{-1}$\\
Reciprocal& $\mu^{-1}=\eta$ & $-\mu^2$ & $2\mu^3$&$\mu^4 V^{-1}$\\
Inverse of the square& $\mu^{-2} = \eta$ & $-\mu^3/2$ & $3\mu^5/4$ & $\mu^6V^{-1}/4$\\
\hline
\end{tabular}
\end{center}
\end{table}
\begin{table}[htb]
\caption{Quantities $q(\mu)$, $V$, $w$ and $\frac{d}{dx}c(\sqrt{V}x+\mu,\phi)$ for some models.}
\begin{center}
\begin{tabular}{ccccccc}
\hline
Model            &$q(\mu)$      & $V$      & $w$                 &  $\frac{d}{dx}c(\sqrt{V}x+\mu,\phi)$\\
\hline
Normal           &$\mu$         & $1$      &  $\mu'^2$           &  $-(x+\mu)\phi$\\
Gamma            &$-1/\mu$      & $\mu^2$  &  $\mu^{-2}\mu'^2$   &  $(\phi-1)/(1+x)$\\
Inverse Gaussian &$-1/(2\mu^2)$ & $\mu^3$  & $\mu^{-3}\mu'^2$    &  $-\frac{3\mu^{3/2}}{2(\mu^{3/2}x+\mu)} + \frac{\phi \mu^{3/2}}{2(\mu^{3/2}x + \mu)^2}$\\
\hline
\end{tabular}
\end{center}
\end{table}
\subsection{Linear models}
For linear models, $\mu_i = \eta_i$, $\mu_i' =1$, $\mu_i'' = 0$,
$w_i = V_i^{-1}$ and $B(\hat{\eta}_i)=0$. Then, $e_i(x) = -V_i^{-1/2}
-\frac{1}{2}V_i^{-1}V_i^{(1)}x$ and $h_i(x) = V_i^{-3/2}V_i^{(1)}+
\frac{3}{4} V_i^{-2}V_i^{(1)^2}x-\frac{1}{2}V_i^{-1}V_i^{(2)}x.$
Hence,
\begin{eqnarray*}
\rho_i(x) &=& V_i^{-1}z_{ii}\,x\left( 1  - \frac{V_i^{-1} V_i^{(1)^2}}{8\phi} + \frac{V_i^{(2)}}{4\phi}+ \frac{V_i^{-1/2}V_i^{(1)}}{2}x\right)\\
&+& \frac{z_{ii}}{2\phi} \left(V_i^{-1} + V_i^{-3/2}V_i^{(1)}\,x + \frac{1}{4} V_i^{-2}V_i^{(1)^2}\,x^2\right)\left\{\phi\sqrt{V_i} q(\mu_i) + \frac{d}{dx}c(\sqrt{V_i}\,x+\mu_i,\phi)\right\}.
\end{eqnarray*}
\subsection{Canonical models}
For canonical models, $\eta_i = \theta_i$, $w_i = V_i$, $\mu_i'=V_i$ and $\mu_i''=V_iV_i^{(1)}$. Further,
$e_i(x) = -V_i^{1/2} -\frac{1}{2}V_i^{(1)}\,x$ and $h_i(x) = \frac{1}{4}(V_i^{(1)^2} - 2V_iV_i^{(2)})\,x$. Hence,
\begin{eqnarray*}
\rho_i(x)\!\!\!\! &=&\!\!\!\! \left(V_i^{1/2}+\frac{V_i^{(1)}}{2}x\right)B(\hat{\eta}_i)+ z_{ii}\left(\frac{V_i^{1/2}V_i^{(1)}}{2\phi} + V_ix +\frac{V_i^{(1)^2}}{8\phi}x + \frac{V_iV_i^{(2)}}{4\phi}x\right) + \frac{V_i^{1/2}V_i^{(1)}}{2} x^2\\
&+& \frac{z_{ii}}{2\phi}\left( V_i + V_i^{1/2}V_i^{(1)}x + \frac{1}{4} V_i^{(1)^2}x^2\right)\left\{\phi\sqrt{V_i}\,q(\mu_i) + \frac{d}{dx} c(\sqrt{V_i}\,x + \mu_i,\phi)\right\}.
\end{eqnarray*}
\subsection{Normal models}
For normal models, $V_i = 1$, $w_i = \mu_i'^2$, $c(x,\phi) = -1/2 \{x^2 \phi + {\rm log}(2\pi/\phi)\}$, $\frac{d}{dx} c(x+\mu,\phi) = -(x+\mu)\phi$, $e_i(x) = -\mu_i'$ and $h_i(x) = -\mu_i''$. We have
$$\rho_i(x) = B(\hat{\eta}_i)\mu_i' + \frac{\mu_i''\,z_{ii}}{2\phi} + \frac{\mu_i'^2z_{ii}}{2}\,x.$$
The normal linear model for which $\mu =\theta =\eta$, $e_i(x) = -1$ and $h_i (x) = 0$ yields
$$\rho_i(x) = z_{ii} \, x/2,$$
and the corrected residuals follow as
$$R_i' = R_i\left(1+\frac{z_{ii}}{2}\right).$$
We can verify that ${\rm Var}(R_i')=1+{\cal O}(n^{-2})$. A check of this expression
can be obtained by considering the simplest case of independent and identically distributed
observations. We have $Z = n^{-1}\,1\,1^T$, $z_{ii} = n^{-1}$ and then
$$R_i' = R_i \left(1+ \frac{1}{2n}\right),$$
which is identical to the result given in the example
discussed by Loynes (1969).
\subsection{Gamma models}
For gamma models, $V_i = \mu_i^2$, $w_i=\mu_i^{-2}\mu_i'^2$, $c(x,\phi) = (\phi-1){\rm log}(x) + \phi{\rm log}(\phi) - {\rm log}\Gamma(\phi)$
and $\frac{d}{dx} c(\mu x+\mu,\phi)=(\phi-1)/(1+x)$. We have $e_i(x)=-\mu_i^{-1}\mu_i'-\mu_i^{-1}\mu_i'x$ and $h_i(x)=-\mu_i^{-1}\mu_i''+ 2\mu_i^{-2}\mu_i'^2-\mu_i^{-1}\mu_i''\,x + 2 \mu_i^{-2}\mu_i'^2\,x.$ Then,
\begin{eqnarray*}
\rho_i(x) = (1+x)\left(\mu_i^{-1}\mu_i'B(\hat{\eta_i}) + \frac{\mu_i^{-1}\mu_i''}{2\phi}z_{ii} - \frac{\mu_i^{-2}\mu_i'^2}{2\phi}z_{ii} + \frac{\mu_i^{-2}\mu_i'^2z_{ii}}{2}x\right).
\end{eqnarray*}
\subsection{Inverse Gaussian models}
For inverse Gaussian models, $V_i = \mu_i^3$, $w_i = \mu_i^{-3}\mu_i'^2$, $c(x,\phi) = (1/2) {\rm log}\{\phi/(2\pi x^3)\} - \phi/(2x)$ and
$\frac{d}{dx} c(\mu^{3/2} x+\mu,\phi) = -\frac{3\mu^{3/2}}{2(\mu^{3/2}x+\mu)} + \frac{\phi \mu^{3/2}}{2(\mu^{3/2}x + \mu)^2}$. Further, $e_i(x) = -\mu_i^{-3/2}\mu_i' - \frac{3}{2}\mu_i^{-1}\mu_i'x$ and $h_i(x) = -\mu_i^{-3/2}\mu_i'' + 3\mu_i^{-5/2}\mu_i'^2 + \frac{15}{4} \mu_i^{-2}\mu_i'^2x - \frac{3}{2} \mu_i^{-1}\mu_i''x$. Then,
\begin{eqnarray*}
\rho_i(x)\!\!\!\! &=&\!\!\!\! \left(\mu_i^{-3/2}\mu_i' + \frac{3\mu_i'}{2\mu_i}x\right)\!\!B(\hat{\eta_i}) +\frac{\mu_i^{3/2}\mu_i''z_{ii}}{2\phi} + \left(\frac{3\mu_i'^2 z_{ii}}{8\phi\mu_i^{2}} + \frac{3\mu_i''z_{ii}}{4\phi\mu_i}\right)\!x +\frac{\mu_i'^2z_{ii}}{\mu_i^{3}}x  + \frac{3\mu_i'^2 z_{ii}}{2\mu_i^{5/2}}\,x^2\\
        &+& \frac{\mu_i^{-3}z_{ii}}{4\phi}\left(\mu_i'^2 + 3\mu_i^{1/2}\mu_i'^2 x+\frac{9\mu_i\mu_i'^2}{4}x^2\right)\left\{-\frac{\phi}{\mu_i^{1/2}} - \frac{3\mu_i^{3/2}}{(\mu_i^{3/2}x+\mu_i)}+\frac{\phi\mu_i^{3/2}}{(\mu_i^{3/2}x+\mu_i)^2}\right\}.
\end{eqnarray*}

\section{Expansion for Cordeiro's adjusted residual}
We now obtain the density function of the adjusted Pearson residuals
proposed by Cordeiro (2004). He gave simple expressions to order
$n^{-1}$ for the mean and variance of the Pearson residual $R_i$ in
GLMs, namely  $E(R_i) = m_i/n + {\cal O}(n^{-2})$ and ${\rm
Var}(R_i) = \sigma^2 + v_i/n + {\cal O}(n^{-2})$, where
$$\frac{m_i}{n}=-\frac{\sigma^2}{2}\gamma_i(I - H)\,J\,z \hbox{~~ and ~~}\frac{v_i}{n}=\frac{\sigma^4}{2}\gamma_i(Q\,H\,J-T)z,$$
$I$ is the identity matrix of order $n$, $H = W^{1/2}X(X^T
WX)^{-1}X^T W^{1/2}$ is the projection matrix, $J,Q$ and $T$ are
diagonal matrices given by $J = {\rm diag}\{V_i^{-1/2}\mu_i''\}$, $Q
={\rm diag}\{V_i^{-1/2} V_i^{(1)}\}$, $T = {\rm diag}\{2\phi w_i+
w_i V_i^{(2)} + V_i^{-1} V_i^{(1)}\mu_i''\}$, $z =
(z_{11},\ldots,z_{nn})^T$ is an $n$-vector with the diagonal
elements of $Z=X(X^TWX)^{-1}X^T$, and $\gamma_i$ was defined in
Section 2. Cordeiro's (2004) adjusted residuals are
\begin {equation}\label{cordeirores}
R_i^{\ast} = \frac{R_i - \hat{m}_i/n}{(\sigma^2 + \hat{v}_i/n)^{1/2}}.
\end{equation}
Expanding $(\sigma^2+\frac{\hat{v}_i}{n})^{-1/2}$ as $\sigma^{-1}(1-
\frac {\hat{v}_i} {2n \sigma^2}+...)$ yields to order $n^{-1}$
$$R_i^{\ast}= \sigma^{-1}\left\{\left(1- \frac {\hat{v}_i} {2n \sigma^2}\right) R_i - \frac {\hat{m}_i}{n}\right\}.$$
Since $\hat{m}_i= m_i+{\cal O}_p(n^{-1/2})$ and $\hat{v}_i= v_i +
{\cal O}_p(n^{-1/2})$, we can write $R_i^\ast$ equivalently to order $n^{-1}$ as
\begin{equation}\label{adjres}
R_i^{\ast}= \sigma^{-1}\left\{R_i-n^{-1}\left(m_i+\frac {v_i R_i}
{2\sigma^2}\right)\right\},
\end{equation}
which implies trivially that $E(R_i^{\ast})=0+{\cal O}(n^{-3/2})$
and ${\rm Var}(R_i^\ast)=1+{\cal O}(n^{-3/2})$. Then, the adjusted
residuals ($\ref{cordeirores}$) have zero mean and unit variance to
order $n^{-1}$.

Let $S_i= \{R_i- n^{-1}(m_i + \frac {v_i R_i} {2\sigma^2})\}$. Since
$R_i={\cal O}_p(1)$, the cumulative distribution function (cdf) of
$S_i$, $F_{S_i}(x)$ say, can be obtained from (\ref{adjres}) to
order $n^{-1}$ following the approach developed by Cordeiro and
Ferrari (1998, Section 2)
\begin{equation}\label{fdaS}
F_{S_i}(x)= F_{R_i}(x)+\frac {1}{n}\left(m_i+ \frac {v_i x}{2\sigma^2}\right)f_{R_i}(x).
\end{equation}
Differentiation of (\ref{fdaS}) with respect to $x$, and replacing
$f_{R_i}(x)$ by its asymptotic expansion in (\ref{loynesglms}),
yields the density of $S_i$ to the same order
\begin{eqnarray}
f_{S_i}(x) &=& f_{\varepsilon_i}(x)-\frac{d}{dx}\{\theta_x^{(i)}f_{\varepsilon_i}(x)\}+\frac{1}{2}\frac{d^2}{dx^2}\{\phi_x^{(i)^2} f_{\varepsilon_i}(x)\}\nonumber\\
&+& \frac{1}{n} \left\{\left(m_i+\frac{v_i x}{2\sigma^2}\right)\frac{df_{\varepsilon_i}(x)}{dx}+ \frac{v_i}{2\sigma^2} f_{\varepsilon_i}(x) \right\}. \label{densS}
\end{eqnarray}
The density function of $R_i^{\ast}$ is $f_{R_i^{\ast}}(x)=\sigma
f_{S_i}(\sigma x)$, where $f_{S_i}(\sigma x)$ comes from
($\ref{densS}$) with $\sigma x$ replacing $x$. The sum of the
second and third terms in ($\ref{densS}$) are expressed as $\frac
{d}{dx}\{\rho_i(x)f_{\varepsilon_i}(x)\}$. Since $m_i/n, v_i/n,
\theta_x^{(i)}$ and $\phi_x^{(i)^2}$ are all quantities of order
${\cal O}(n^{-1})$, the terms on the right hand side of
($\ref{densS}$), except $f_{\varepsilon_i}(x)$, are of this order
and then the densities $f_{R_i^{\ast}}(x)$ and
$f_{\varepsilon_i}(x)$ differ by terms of order ${\cal O}(n^{-1})$.
However, we showed in Section 3, that the densities $f_{R_i'}(x)$
and $f_{\varepsilon_i}(x)$ are equal to this order. Thus, the
distribution of the corrected residuals $R_i'$, even in small
samples, is closer to the distribution of the true Pearson residuals
than the distribution of the adjusted residuals
$R_i^\ast$.

A simple expansion for the density $f_{R_i^{\ast}}(x)$ of the adjusted residuals $R_i^{\ast}$
to order $n^{-1}$ for the normal model with any link function is given by
\begin{eqnarray*}
f_{R_i^{\ast}}(x)=\frac{e^{-\frac{x^2}{2}}}{\sqrt{2\pi}}\left(1+a_0-a_1 x- a_2 x^2\right),
\end{eqnarray*}
where the constant terms
$$a_0=\frac{3\mu_i'^2z_{ii}}{2} +\frac{v_i}{2n\sigma^2},\,a_1=\frac{m_i}{\sigma n}-\frac{\mu_i'B(\hat{\eta_i})}{\sigma}-\frac{\sigma\mu_i''z_{ii}}{2} \hbox{~and~}a_2=\frac{v_i}{2n\sigma^2}-\frac{3\mu_i'^2z_{ii}}{2}$$ 
that depend on the model are all of order ${\cal O}(n^{-1})$. 

\section{Simulation results}

We present some simulation results for studying the finite-sample 
distributions of the Pearson $R_i$, corrected $R_i'$, adjusted $R_i^\ast$ 
and the true $\varepsilon_i$ residual. We use a gamma model with log link
$${\rm log}\,\mu = \beta_0 + \beta_1 x_1 + \beta_2 x_2,$$
where the true values of the parameters were taken as $\beta_0 =
1/2$, $\beta_1 = 1$, $\beta_2=-1$ and $\phi=4$. The explanatory
variables $x_1$ and $x_2$ were generated from the uniform U$(0,1)$
distribution for $n=20$ and their values were held constant
throughout the simulations. The number of Monte Carlo replications
was set at $10,000$ and all simulations were performed using the
statistical software {\tt R}.

In each of the $10,000$ replications, we fitted the model and computed the
MLE $\hat\beta$ and fitted mean $\hat{\mu}$, the Pearson residuals $R_i$, the corrected
function $\rho(\cdot)$ and the corrected residuals $R_i'$. 
Further, we calculated their expected values and variances from the expressions given 
by Cordeiro (2004) to obtain the adjusted residuals $R_i^\ast$. Finally, we calculated 
the true residuals $\varepsilon_i$. Tables 4 and 5 give the sample means, variances, skewness 
and kurtosis of the residuals $R_i$, $R_i'$, $R_i^\ast$ and $\varepsilon_i$, respectively, out of $10,000$
values. The corrected residuals $R_i'$ should agree with the true Pearson residuals
rather than to the normal distribution. A good agreement with the normal distribution happens
when these figures are, on average, close to 0,1,0 and 3, respectively.

The figures in Tables 4 and 5 show that the distribution of all residuals for the gamma model are
positively skewed. All four cumulants of the corrected Pearson residuals $R_i'$ are generally closer to the
corresponding cumulants of the true residuals $\varepsilon_i$ than those of the other residuals. The adjusted
residuals $R_i^\ast$ have cumulants much closer to the cumulants of a standard normal distribution
as claimed by Cordeiro (2004). Further, the distribution of the corrected residuals
is generally closer to the distribution of the true residuals than the distribution of
the Pearson residuals. In short, the correction $\rho(.)$ appears to be effective even when the sample
size is small.

In Table 6 we give the values of the Kolmogorov-Smirnov (K-S) and Anderson-Darling (A-D) (see, for instance,
Anderson and Darling, 1952; Thode, 2002, Section 5.1.4) distances between the empirical distribution of
each set of the $10,000$ uncorrected $R_i$ and corrected $R_i'$ residuals for
$i=1,\ldots,20$, and the estimated distribution of the true re\-si\-duals. The estimated 
distribution here is the shifted gamma distribution with dispersion parameter $\phi$ taken to be the sample average of the estimated dispersion
parameters at each step of the Monte Carlo simulation. In Table 7, we follow the same procedure for Table 6,
but we now examine if the uncorrected $R_i$ and corrected $R_i'$ residuals follow the empirical distribution 
of the true residual $\varepsilon_i$. We then calculated both K-S and A-D distances between the 
empirical distributions of both (uncorrected and corrected) residuals and the empirical distribution 
of the true residuals $\varepsilon_i$.

We see from Tables 6 and 7 that the distribution of the corrected residuals is closer to the distribution
of the true residuals than the distribution of the uncorrected residuals. Furthermore, the distances 
for the corrected residuals are substantially smaller than the distances for the uncorrected ones.
These facts show that, when the model is well-specified, our correction works very well for the set of 
the corrected residuals.
\begin{table}[htb]
\caption{Mean and variance of uncorrected, corrected, adjusted and true residuals.}
\begin{center}
\begin{tabular}{lrrrrrrrrrrrrr}
\hline
&&\multicolumn{4}{c}{Mean}&&&&\multicolumn{4}{c}{Variance}\\
&&\multicolumn{4}{c}{--------------------------------------}&&&&\multicolumn{4}{c}{-----------------------------------}\\
$i$&&\multicolumn{1}{c}{$R_i$}&\multicolumn{1}{c}{$R_i'$}&\multicolumn{1}{c}{$R_i^\ast$}&\multicolumn{1}{c}{$\varepsilon_i$}&&&&\multicolumn{1}{c}{$R_i$}&\multicolumn{1}{c}{$R_i'$}&\multicolumn{1}{c}{$R_i^\ast$}&\multicolumn{1}{c}{$\varepsilon_i$}\\
\hline
1&   &  0.013 &  0.006 & 0.011 & 0.004&& &  & 0.234& 0.255  &1.059& 0.257\\
2&   & -0.010 & -0.006 & 0.007 & 0.001&& &  & 0.183& 0.232  &1.112& 0.255\\
3&   &  0.002 & -0.002 &-0.004 &-0.002&& &  & 0.220& 0.248  &1.040& 0.254\\
4&   &  0.006 &  0.003 & 0.010 & 0.005&& &  & 0.208& 0.241  &1.051& 0.253\\
5&   &  0.015 &  0.004 & 0.005 & 0.002&& &  & 0.237& 0.247  &1.006& 0.249\\
6&   & -0.003 & -0.005 & 0.003 &-0.001&& &  & 0.188& 0.229  &1.043& 0.245\\
7&   & -0.002 & -0.006 &-0.008 &-0.005&& &  & 0.201& 0.237  &1.038& 0.244\\
8&   & -0.012 & -0.009 & 0.001 &-0.001&& &  & 0.180& 0.230  &1.107& 0.258\\
9&   &  0.000 & -0.001 & 0.008 & 0.000&& &  & 0.201& 0.244  &1.087& 0.253\\
10&  & -0.005 & -0.010 &-0.014 &-0.010&& &  & 0.207& 0.235  &0.999& 0.236\\
11&  & -0.000 &  0.001 & 0.010 & 0.002&& &  & 0.201& 0.244  &1.079& 0.254\\
12&  &  0.010 & -0.001 &-0.006 &-0.000&& &  & 0.243& 0.252  &1.022& 0.259\\
13&  & -0.009 & -0.012 &-0.016 &-0.009&& &  & 0.199& 0.230  &1.002& 0.239\\
14&  & -0.003 & -0.001 & 0.019 & 0.005&& &  & 0.176& 0.227  &1.116& 0.248\\
15&  &  0.012 &  0.005 & 0.010 & 0.004&& &  & 0.221& 0.243  &1.017& 0.252\\
16&  & -0.017 & -0.017 &-0.014 &-0.007&& &  & 0.174& 0.225  &1.105& 0.249\\
17&  &  0.004 & -0.004 &-0.009 &-0.004&& &  & 0.221& 0.241  &1.022& 0.246\\
18&  &  0.001 & -0.004 &-0.005 &-0.004&& &  & 0.214& 0.240  &1.020& 0.246\\
19&  &  0.000 & -0.004 &-0.006 &-0.002&& &  & 0.215& 0.239  &1.019& 0.249\\
20&  & -0.003 & -0.004 &-0.003 &-0.002&& &  & 0.196& 0.230  &1.008& 0.240\\
\hline
\end{tabular}
\end{center}
\end{table}
\begin{table}[htb]
\caption{Skewness and kurtosis of uncorrected, corrected, adjusted and true residuals.}
\begin{center}
\begin{tabular}{lrrrrrrrrrrrrr}
\hline
&&\multicolumn{4}{c}{Skewness}&&&&\multicolumn{4}{c}{Kurtosis}\\
&&\multicolumn{4}{c}{-----------------------------------}&&&&\multicolumn{4}{c}{-----------------------------------}\\
$i$&&\multicolumn{1}{c}{$R_i$}&\multicolumn{1}{c}{$R_i'$}&\multicolumn{1}{c}{$R_i^\ast$}&\multicolumn{1}{c}{$\varepsilon_i$}&&&&\multicolumn{1}{c}{$R_i$}&\multicolumn{1}{c}{$R_i'$}&\multicolumn{1}{c}{$R_i^\ast$}&\multicolumn{1}{c}{$\varepsilon_i$}\\
\hline
1&   & 0.837& 0.943 &0.626& 1.005&& &  &3.798& 4.105  &  2.967 & 4.468\\
2&   & 0.586& 0.822 &0.494& 0.986&& &  &3.205& 3.780  &  2.805 & 4.399\\
3&   & 0.824& 0.973 &0.605& 1.080&& &  &3.898& 4.350  &  3.020 & 4.859\\
4&   & 0.703& 0.863 &0.550& 0.979&& &  &3.417& 3.825  &  2.882 & 4.395\\
5&   & 0.876& 0.942 &0.626& 0.964&& &  &4.040& 4.232  &  3.012 & 4.275\\
6&   & 0.628& 0.829 &0.523& 0.987&& &  &3.278& 3.772  &  2.823 & 4.387\\
7&   & 0.715& 0.901 &0.548& 0.960&& &  &3.548& 4.052  &  2.920 & 4.317\\
8&   & 0.611& 0.864 &0.500& 1.068&& &  &3.318& 3.984  &  2.865 & 4.813\\
9&   & 0.711& 0.923 &0.557& 1.017&& &  &3.561& 4.162  &  2.911 & 4.667\\
10&  & 0.809& 0.965 &0.628& 1.018&& &  &3.904& 4.387  &  3.061 & 4.811\\
11&  & 0.727& 0.936 &0.556& 1.018&& &  &3.590& 4.176  &  2.920 & 4.532\\
12&  & 0.939& 1.001 &0.659& 1.077&& &  &4.361& 4.560  &  3.076 & 4.929\\
13&  & 0.746& 0.907 &0.603& 0.938&& &  &3.607& 4.052  &  3.006 & 4.106\\
14&  & 0.553& 0.801 &0.474& 0.939&& &  &3.150& 3.709  &  2.820 & 4.254\\
15&  & 0.808& 0.928 &0.606& 1.048&& &  &3.813& 4.154  &  3.033 & 4.737\\
16&  & 0.593& 0.851 &0.506& 1.006&& &  &3.246& 3.833  &  2.859 & 4.510\\
17&  & 0.793& 0.910 &0.606& 0.958&& &  &3.727& 4.058  &  2.994 & 4.202\\
18&  & 0.783& 0.923 &0.610& 0.992&& &  &3.686& 4.078  &  2.977 & 4.411\\
19&  & 0.776& 0.904 &0.603& 0.963&& &  &3.687& 4.060  &  2.993 & 4.292\\
20&  & 0.715& 0.888 &0.569& 0.963&& &  &3.532& 4.004  &  2.960 & 4.346\\
\hline
\end{tabular}
\end{center}
\end{table}

\begin{table}[htb]
\caption{One-sample K-S and A-D statistics for uncorrected and corrected residuals.}
\begin{center}
\begin{tabular}{lcrcr}
\hline
$i$&K-S stat. for $R_i$&A-D stat. for $R_i$&K-S stat. for $R_i'$&A-D stat. for $R_i'$\\
\hline
All&$0.0232 $&$300.42$&$0.0036$&$6.7641$\\
1  &$0.0230 $&$ 7.5944 $&$0.0103$&$1.7875$\\
2  &$0.0317 $&$30.7031 $&$0.0074$&$1.2504$\\
3  &$0.0208 $&$ 7.9810 $&$0.0077$&$1.0710$\\
4  &$0.0287 $&$17.5666 $&$0.0100$&$1.2283$\\
5  &$0.0216 $&$ 8.9498 $&$0.0098$&$1.1761$\\
6  &$0.0307 $&$28.3464 $&$0.0074$&$0.8353$\\
7  &$0.0273 $&$17.8230 $&$0.0088$&$1.0719$\\
8  &$0.0311 $&$34.7206 $&$0.0109$&$1.7666$\\
9  &$0.0277 $&$19.3796 $&$0.0081$&$0.9986$\\
10 &$0.0244 $&$12.6919 $&$0.0123$&$1.9530$\\
11 &$0.0306 $&$19.5087 $&$0.0089$&$0.6709$\\
12 &$0.0167 $&$ 3.6271 $&$0.0106$&$2.0631$\\
13 &$0.0208 $&$15.1071 $&$0.0107$&$2.2356$\\
14 &$0.0401 $&$49.8411 $&$0.0117$&$1.6905$\\
15 &$0.0277 $&$16.1354 $&$0.0150$&$1.8709$\\
16 &$0.0360 $&$43.4746 $&$0.0155$&$2.5022$\\
17 &$0.0186 $&$ 7.3023 $&$0.0082$&$1.0216$\\
18 &$0.0235 $&$ 9.8595 $&$0.0068$&$0.6481$\\
19 &$0.0172 $&$ 7.6480 $&$0.0085$&$0.8816$\\
20 &$0.0282 $&$21.7356 $&$0.0072$&$0.7146$\\
\hline
\end{tabular}
\end{center}
\end{table}

\begin{table}[htb]
\caption{Two-sample K-S and A-D statistics for uncorrected and corrected residuals.}
\begin{center}
\begin{tabular}{lcrcr}
\hline
$i$&K-S stat. for $R_i$&A-D stat. for $R_i$&K-S stat. for $R_i'$&A-D stat. for $R_i'$\\
\hline
All&$0.0283$&$444.193$&$0.0041$&$ 9.5703$\\
1  &$0.0246$&$10.5690$&$0.0086$&$0.5191$\\
2  &$0.0356$&$44.3716$&$0.0125$&$1.1887$\\
3  &$0.0273$&$15.8880$&$0.0100$&$0.8209$\\
4  &$0.0331$&$22.3655$&$0.0079$&$0.9083$\\
5  &$0.0227$&$9.41018$&$0.0070$&$0.4727$\\
6  &$0.0355$&$33.9614$&$0.0094$&$1.5002$\\
7  &$0.0348$&$23.8043$&$0.0077$&$0.5297$\\
8  &$0.0394$&$54.3072$&$0.0126$&$1.3377$\\
9  &$0.0325$&$27.3448$&$0.0071$&$0.5715$\\
10 &$0.0270$&$14.1363$&$0.0065$&$0.2252$\\
11 &$0.0336$&$28.1285$&$0.0100$&$0.8706$\\
12 &$0.0218$&$9.17441$&$0.0102$&$0.4995$\\
13 &$0.0342$&$22.1684$&$0.0112$&$1.2824$\\
14 &$0.0426$&$56.0126$&$0.0132$&$2.5277$\\
15 &$0.0281$&$15.7758$&$0.0121$&$1.1146$\\
16 &$0.0444$&$57.0837$&$0.0109$&$1.9487$\\
17 &$0.0255$&$14.0805$&$0.0081$&$0.5073$\\
18 &$0.0310$&$16.4867$&$0.0089$&$0.4553$\\
19 &$0.0282$&$16.7720$&$0.0094$&$1.0549$\\
20 &$0.0303$&$22.7455$&$0.0087$&$0.8158$\\
\hline
\end{tabular}
\end{center}
\end{table}

We conclude the study providing an application of the corrected residuals to assess the adequacy of
the above gamma model. We could expect that under a well-specified model, the distribution of the corrected
residuals will follow approximately the distribution of the true residuals. However, even though it
is common to compare the distribution of the Pearson residuals with the normal distribution, it is not
clear that this approximation should be good in small samples. Therefore, we compare the empirical distribution
of the corrected residuals with the distribution of the true residuals and the distribution of the uncorrected
residuals with the normal distribution. For doing this, we use a QQPlot which displays a quantile-quantile plot
of the sample quantiles of the corrected and uncorrected residuals versus theoretical quantiles from the estimated
distribution of the true residuals and the normal distribution with mean zero and variance $\hat\phi^{-1}$, respectively.
If the distribution of the corrected residuals is well approximated by the distribution of the true residuals,
the plot will be close to linear. Therefore, we expect that a QQPlot of the Studentized corrected residuals
versus the estimated distribution of the true residuals should be closer to the diagonal line than that
QQPlot of the uncorrected residuals against the normal $N(0,\hat\phi^{-1})$ distribution. Moreover, we also consider
the QQPlot of the adjusted residuals suggested by Cordeiro (2004) against the theoretical quantiles of a standard
normal distribution.

Figure 1 gives two QQPlots, one for the vector of the $10,000$ ordered uncorrected residuals and other for the vector
of the $10,000$ ordered corrected residuals. These fi\-gu\-res show that even for a well-specified model, the plot for
the uncorrected residuals is very distant from the diagonal line when compared with the plot for the corrected
residuals. The adjusted residuals given in Figure 2 provides an improvement in regard to the uncorrected residuals,
but the plot is also distant from the diagonal line when compared to the corrected residuals. Therefore, the
corrected residuals have a good behavior that leads to the right conclusion, i.e., that the model is 
well-specified. We thus recommend the corrected residuals to build up QQPlots.
\begin{figure}[htb]
\begin{center}
\includegraphics[scale=0.55]{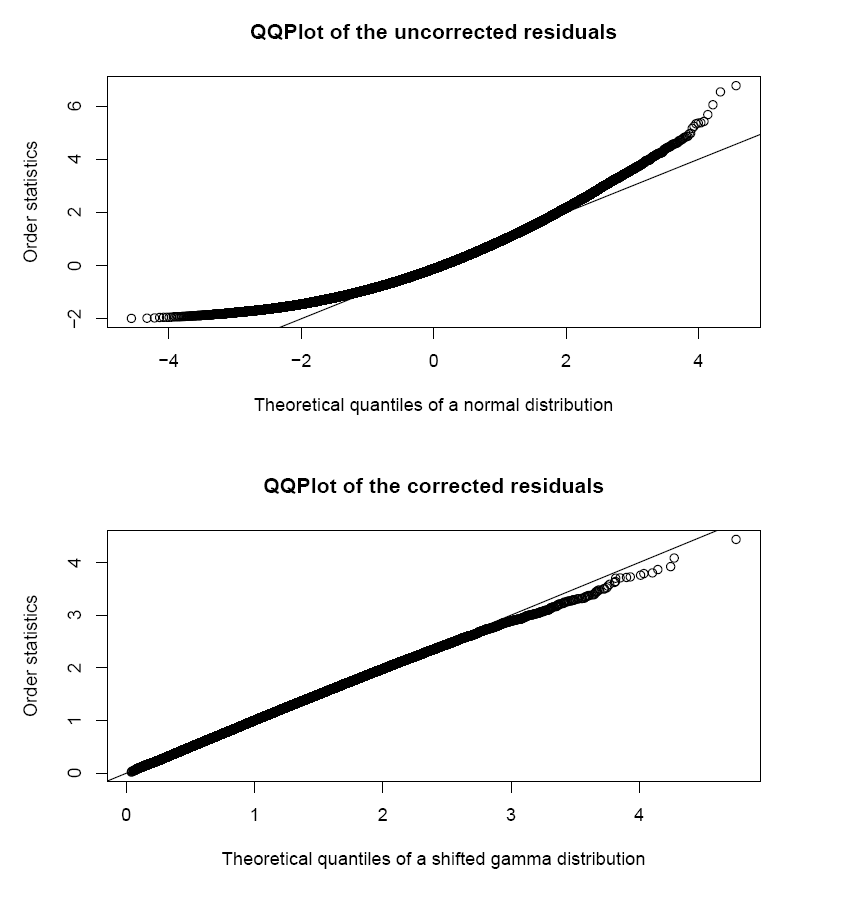}
\caption{QQPlots for the Pearson and corrected residuals}
\end{center}
\end{figure}
\begin{figure}[htb]
\begin{center}
\includegraphics[scale=0.55]{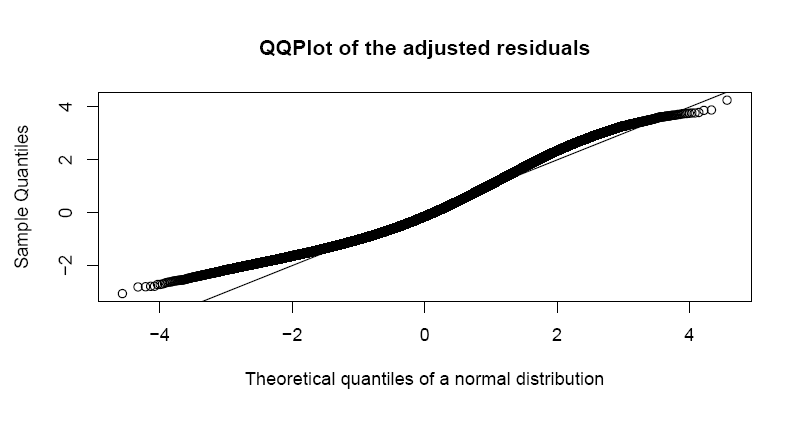}
\caption{QQPlot for the adjusted residuals}
\end{center}
\end{figure}

\newpage
\section{Conclusion}

Using the results given in Loynes (1969), we calculate the
${\cal O}\left(n^{-1}\right)$ distribution of the Pearson
residuals in GLMs (see, for instance, McCullagh and Nelder, 
1989). It is important to mention that
the distribution of residuals in regression models are
typically unknown, and therefore all inference regarding
these residuals are done by asymptotic assumptions which may 
not hold in small or moderate sample sizes. Then we can use this 
knowledge to define corrected Pearson residuals in these models in such a way 
that the corrected residuals will have, to order ${\cal O}\left(n^{-1}\right)$, 
the same distribution of the true Pearson residuals, which is known. The
corrected residuals have practical applicability for all continuous GLMs.
We simulate a gamma model with log link to conclude the superiority of the corrected Pearson residuals 
$R_i'$ over the uncorrected residuals $R_i$ and also over the adjusted residuals suggested
by Cordeiro (2004) with regard to the approximation to the reference distribution, 
which for the corrected and uncorrected residuals was the distribution of the true residuals and 
for the adjusted residuals was the standard normal distribution. The paper is concluded with an 
application of the corrected residuals to assess the adequacy of the model.

\section*{Appendix}
Suppose we write the residual $R$ in terms of the true residual
$\varepsilon$ as $R = \varepsilon + \delta$, where $\varepsilon$ and
$\delta$ are absolutely continuous random variables with respect to
Lebesgue measure and $\delta$ is of order ${\cal O}_p(n^{-1})$. Our goal is to
define a corrected residual $R'$ having the same density of $\varepsilon$
to order $n^{-1}$. Initially, we have
$$E(e^{isR}) = E\{e^{is\varepsilon}E(e^{is\delta}\mid\varepsilon)\}\hbox{~and~}\left.\frac{\partial^k}{\partial s^k} E(e^{is\delta}\mid\varepsilon)\right\vert_{s=0} = i^k E(\delta^k \mid \varepsilon).$$
Expanding $E(e^{is\delta}\mid\varepsilon)$ in a Taylor series around $s=0$ gives
$$E(e^{i s \delta}\mid\varepsilon) = 1 + (is)E(\delta\mid\varepsilon) + \frac{(is)^2}{2}E(\delta^2\mid\varepsilon) + \cdots.$$
Let $\theta_x = E(\delta\mid\varepsilon=x)$ and $\phi_x^2 = {\rm Var}(\delta\mid\varepsilon=x)$. Thus,
\begin{equation}\label{expect}
E\{e^{is\varepsilon}E(e^{i s \delta}\mid\varepsilon)\} = \int_{-\infty}^\infty e^{isx} \left\{1 + (is)\theta_x + \frac{(is)^2}{2}(\phi_x^2+\theta_x^2) + \cdots\right\} f_\varepsilon(x)dx,
\end{equation}
where $f_\varepsilon(\cdot)$ is the density function of $\varepsilon$.
By using formulae (25) and (26) from Cox and Snell (1968) with $\varepsilon=0$,
it is possible to conclude that $E(\delta)$ and Var$(\delta)$ (and thus $E(\delta^2)$)
are of order ${\cal O}(n^{-1})$ and, in the same way, that the higher moments
of $\delta$ are of order $o(n^{-1})$. In a similar manner, we can show that $E(\delta\mid\varepsilon = x)$ and Var$(\delta\mid\varepsilon =x)$ are also of order ${\cal O}(n^{-1})$, and that the higher-order conditional moments are of
order $o(n^{-1})$. Then, we can rewrite equation (\ref{expect}) as
\begin{equation}\label{expect2}
E\{e^{is\varepsilon}E(e^{is\delta}\mid\varepsilon)\} = \int_{-\infty}^\infty e^{isx} \left\{1 + (is)\theta_x + \frac{(is)^2}{2}\phi_x^2\right\}\,f_\varepsilon(x)dx + o(n^{-1}).
\end{equation}
Note that we can express the integral on the right side of (\ref{expect2}) as a sum of three integrals. Then, 
integration by parts, one time for the integral containing $\theta_x$ on the integrand and two times for the
integral containing $\phi_x^2$ on the integrand, yields the following formula
\begin{equation}\label{expect3}
E(e^{isR}) = \int_{-\infty}^\infty e^{isx}\left[f_\varepsilon(x) - \frac{d\{f_\varepsilon(x)\theta_x\}}{dx} + \frac{1}{2}
\frac{d^2\{f_\varepsilon(x)\phi_x^2\}}{dx^2}\right]dx + o(n^{-1}).
\end{equation}
The uniqueness theorem for characteristic functions yields the density of $R$
to order $n^{-1}$
\begin{equation}\label{densloynes}
f_{R}(x) = f_{\varepsilon}(x) - \frac{d \{f_{\varepsilon}(x) \theta_x\}}{dx}
+ \frac{1}{2} \frac{d^2 \{f_{\varepsilon}(x) \phi_x\}}{dx^2}+ o({n^{-1}}).
\end{equation}
Equation ($\ref{densloynes}$) is identical to formula (5) in Loynes (1969).

Further, we now define corrected residuals of the form
$R' = R + \rho (R),$ where $\rho(\cdot)$ is a function of order
${\cal O}(n^{-1})$ used to recover the distribution of $\varepsilon$.
We may proceed as above, noting that $E\{\rho(R)\mid R = x\} = \rho(x)$,
to obtain the density of $R'$ to order $n^{-1}$
$$f_{R'}(x) = f_R(x) - \frac{d}{dx}\{\rho(x)f_R(x)\}.$$
Since the quantities $\rho(x), \theta_x$ and $\phi_x^2$ are all of
order ${\cal O}(n^{-1})$, we have that $\frac{d}{dx}\{\rho(x)f_R(x)\} = \frac{d}{dx}\{\rho(x)f_\varepsilon(x)\}$ to this order.
Therefore, the densities of $R$ and $\varepsilon$ will be the same to order
$n^{-1}$ if
$$\frac{d}{dx}\{\rho(x)f_{\varepsilon}(x)\} = -\frac{d}{dx}\{f_{\varepsilon}(x) \theta_x\}
+ \frac{1}{2} \frac{d^2}{dx^2} \{f_{\varepsilon}(x) \phi_x\}.$$
Integration gives
\begin{equation}\label{eqloynes}
\rho(x) = - \theta_x + \frac{1}{2 f_{\varepsilon}(x)} \frac{d}{dx} \{f_{\varepsilon}(x) \phi_x\}.
\end{equation}
Equation (\ref{eqloynes}) is identical to equation (6) given by
Loynes (1969) and it is clear from the proof that the support of $\varepsilon$ does not
need to be the entire line and we can have proper intervals as support. We should note that the assumptions needed can be made weaker if
we require that an expansion of the Taylor polynomial of order two with a
remainder term (for instance, Lagrange remainder) can be done instead
of the complete series.

We could also prove Loynes' (1969) results by using the equivalence of (3c) and (4c), together with (5) and (6) of Cox and Reid (1987) and appropriate
regularity conditions. The idea to this approach is as follows: consider in equation (3c) of Cox and Reid (1987)
$X_0 = \varepsilon$, $X_1 = n^{1/2}\delta$ and $X_2=0$. This means that we are writing $Y_n$ as $Y_n = \varepsilon+\delta+{\cal O}_p(n^{-3/2})$,
where $\varepsilon$ and $\delta$ are of orders ${\cal O}_p(1)$ and ${\cal O}_p(n^{-1})$, respectively. Then, from (4c), (5) and 
(6) of Cox and Reid (1987), we can write de cdf of $Y_n$ as
$$G_n(y) = F_0(y) - E(\delta\mid\varepsilon=y)f_0(y) + \frac{1}{2}\frac{\partial}{\partial y}\{E(\delta^2\mid\varepsilon=x)f_0(y)\} + {\cal O}(n^{-3/2}),$$
where $F_0(\cdot)$ and $f_0(\cdot)$ are the cdf and pdf of $\varepsilon$, respectively. The expression above 
implies equation (\ref{densloynes}). We can also obtain the expansion for $R + \rho(R)$ from the equivalence of (3c) 
and (4c) of Cox and Reid (1987) by setting $X_0=R, X_1=0$ and $X_2 = \rho(R)$. The rest of the proof is 
identical to the one given before. Note also, that for this proof $\varepsilon$ does not need to have a 
support in the entire line since this is not an assumption in the usual regularity conditions.

\end{document}